\documentclass[prd,aps,floatfix,nofootinbib,preprint,tightenlines ]{revtex4}
\usepackage{dcolumn}
\usepackage{amsmath}
\usepackage{latexsym}
\usepackage{graphicx}
\usepackage{bm}

\def\svev#1{\left\langle #1\right\rangle}       

\def\Tr{{\rm Tr}\,}

\def\Re{{\rm Re\,}}
\def\Im{{\rm Im\,}}

\newcommand{\bee}{\begin{equation}}
\newcommand{\ee}{\end{equation}}
\newcommand{\beea}{\begin{eqnarray}}
\newcommand{\eea}{\end{eqnarray}}

\begin{document}
\title{
Finite-size scaling tests for SU(3) lattice gauge theory with color sextet fermions}
\author{Thomas DeGrand}
\email{thomas.degrand@colorado.edu}
\affiliation{Department of Physics,
University of Colorado, Boulder, CO 80309, USA \\
The Niels Bohr International Academy, The Niels Bohr Institute,
Blegdamsvej 17, DK-2100 Copenhagen \O, Denmark}

\begin{abstract}
The observed slow running of the gauge coupling in SU(3) lattice gauge theory with two
flavors of
 color sextet fermions naturally suggests it is a theory with one relevant
coupling, the fermion mass, and that at zero mass correlation functions decay
algebraically.
I perform a finite-size scaling study on simulation data at two values of
the bare gauge coupling with this assumption and observe 
a common exponent for the scaling of the correlation length with the fermion mass,
$y_m \sim 1.5$. An analysis of the scaling of valence Dirac eigenvalues
at one of these bare couplings produces a similar number.
\end{abstract}

\maketitle

\section{Introduction and theoretical background}
Recently, many researchers
\cite{Damgaard:1997ut,Heller:1997vh,
Catterall:2007yx,Appelquist:2007hu,Shamir:2008pb,Deuzeman:2008sc,DelDebbio:2008zf,Catterall:2008qk,
Fodor:2008hm,DelDebbio:2008tv,DeGrand:2008kx,
Hietanen:2008mr,Fleming:2008gy,Appelquist:2009ty,Hietanen:2009az,Deuzeman:2009mh,Fodor:2009wk,DelDebbio:2009fd,Pica:2009hc}
have begun to use lattice methods to
 study field theories which might be candidates for strongly coupled beyond - Standard Model
new physics \cite{Hill:2002ap}.
 These models typically involve gauge fields and a large number of fermion degrees of freedom,
either many flavors of fundamental representation fermions
(where the discussion goes back to Refs.~\cite{Caswell:1974gg,Banks:1981nn})
 or a smaller number of flavors of 
higher-dimensional representation fermions
(strongly emphasized by Refs.~\cite{Sannino:2004qp,Dietrich:2006cm,Ryttov:2007sr}).

The usual description of renormalization for a gauge theory coupled to massless fermions
classifies the possibilities for its behavior according to how
the gauge coupling flows under rescaling to the infrared: it could flow
to strong coupling in the case of a confining theory, or to zero, for a trivial theory, or
 to a fixed point at some nonzero value, $g^{2*}$. 
 The latter case is referred to as an
infrared-attractive fixed point (IRFP) theory.
This is a phase with no confinement,
 no chiral symmetry breaking, and algebraic decay of correlation functions.
Evidence has been presented that several such theories exist:
$SU(3)$ gauge theory with $N_f=12$ flavors of fundamental fermions
 \cite{Appelquist:2007hu,Appelquist:2009ty},
$SU(3)$ gauge theory with $N_f=2$ flavors of sextet fermions (the subject of this paper)
\cite{Shamir:2008pb}
and $SU(2)$ gauge theory with $N_f=2$ flavors of adjoint fermions\cite{Hietanen:2009az}.

While for confining theories the gauge coupling is relevant (the Gaussian fixed point $g=0$ is
unstable), that is not the case for IRFP theories.
The  distance of the bare gauge coupling from its fixed point value
$(g^2-g^{2*})$ is an irrelevant coupling. The critical surface encompasses a wide range
of values of bare gauge couplings.
The fermion mass is a relevant coupling of an IRFP theory,
since it  must be fine-tuned at the UV scale to reach the critical surface (i.e. $m_q=0$).
The situation is completely equivalent to that of an order-disorder transition in a magnetic
 system. The only difference is that the relevant direction
is parameterized by the quark mass $m_q$, rather than by the reduced temperature
 $t=(T-T_c)/T_c$ of the magnet. (A closer analogy is to a system which has been fine
 tuned to its Curie point
and then placed in an external magnetic field. The external field breaks the underlying global 
symmetry just as a quark mass explicitly breaks chiral symmetry.) \cite{DeGrand:2009mt}

The framework to describe the physics of these systems is standard.
It involves a set of scaling operators which evolve independently and multiplicatively
and whose renormalization group (RG) equations have linear zeroes;
the RG equation for the change in the $i$th coupling, under a scale change by a factor $s$,
linearized around its fixed-point value $g_i^*$, is
\bee
s\frac{ dg_i}{ds} = y_i(g_i-g_i^*) .
\ee 
The mass is the relevant coupling and  its exponent will be labeled $y_m$.
Tuning the mass to zero causes the correlation length to diverge algebraically,
\bee
\xi \sim m_q^{-\frac{1}{y_m}}.
\label{eq:corrlen}
\ee
The exponent $y_m$ is related to the evolution of the condensate in 
beyond-Standard Model phenomenology. Under a change of scale from $s_1$ to $s_2$,
 the condensate runs  as
\bee
\langle \bar\psi\psi \rangle_{s_2} = \langle \bar\psi\psi \rangle_{s_1}
\exp\int_{s_1}^{s_2}\frac{d\mu}{\mu}d(\mu)
\label{eq:gammaruns}
\ee
where $d$ is its scaling dimension. ($d=3+\gamma_{\bar\psi\psi}$ where
$\gamma_{\bar\psi\psi}$ is its anomalous dimension).
Because the combination $m\bar \psi \psi$ is an RG invariant, $y_m = 4-d$.

Thus $y_m$ is an interesting object\cite{Luty:2008vs}.
 The subject of this paper is a calculation of $y_m$
for $SU(3)$ gauge fields coupled to $N_f=2$ flavors of fermions in the sextet (color symmetric)
representation.

Most recent lattice work on candidate theories is devoted to answering the question
of whether or not the gauge coupling runs to a fixed point.
The analysis in this paper simply assumes that the gauge coupling runs very slowly.
What happens in that case can be seen from an elaboration of  Eq.~\ref{eq:gammaruns}.
 Suppose that we have
a correlation function measured on a momentum scale which is large compared to
other possible scales in the theory. Then, textbooks tell us that a generic correlation
function (with engineering dimension $d_n$ and
 associated anomalous dimension $\gamma$) scales as
\bee
\Gamma(sp) = s^{d_n} \Gamma(p) \exp \int_1^s \frac{dt}{t}\gamma(g(t)).
\label{eq:gammaruns2}
\ee
When the integral is dominated by scales where the coupling is given by its fixed point value,
then $\Gamma(k) \sim k^{d_n+\gamma}$
where $\gamma=\gamma(g^*)$. The correlator has power law behavior.
(This behavior is modified at short distances by non-scaling terms in the action.)

If we have a real fixed point, then  the critical exponents (in this case, $y_m$)
 will not depend on the value of the bare coupling.  However,
imagine that we do not have a fixed point. Then the integral in Eq.~\ref{eq:gammaruns2}
will depend on $s$. But if the change in the coupling over the range of the integral is small,
$\gamma(g(t))$ will remain unchanged, and one will again observe a power law behavior for correlators.
Of course, the value of the exponent  $\gamma$ would change as the bare coupling were
varied. Whether one is seeing a real exponent (a constant $y_m$) or an effective one (that is, one is
mapping out $y_m(g)$) can be tested by determining  $y_m$ for several values of the bare gauge coupling.

This discussion would be redundant if we knew from other sources
 whether the theory was in the IRFP phase or not.
It is necessary to make it because this part of the story is not complete.
It is quite clear that, in my system's weak coupling phase, 
 observables (correlation lengths) depend quite strongly on the
quark mass and only weakly on the gauge coupling. The correlation length saturates
at a value proportional to the size of the system as the quark mass is tuned to zero.

And in the weak coupling phase, the running coupling does run slowly.
Last year Svetitsky, Shamir and I \cite{Shamir:2008pb}
presented evidence that $N_f=2$ sextet fermions
and $SU(3)$ gauge fields had an IRFP. Those simulations were done with a different bare action
than the one used here. We are repeating and extending our calculation of the running
coupling using the present bare action. That analysis is at present
 incomplete \cite{BQStalk}.
At this point in time, we know that throughout the weak coupling phase
 the running coupling (defined through the Schrodinger
Functional (SF) scheme) runs more slowly than two-loop perturbation theory predicts.
This itself is slow running compared to the familiar case of $N_f=2$ fundamental
representation QCD. That this should be so is obvious from the beta function,
but the numerical values are worth mentioning. 
At the two bare couplings $\beta=6/g^2$ where I will claim a measurement of $y_m$,
  the SF coupling
(measured on $6^4$ volumes) is about $g_{SF}^2=2.5$ at $\beta=5.2$ and about 3.4
at $\beta=4.8$. (Parameter sets will be given below.) 
The change of $g_{SF}^2$ in a scale factor of 2
from  integrating the two-loop perturbative beta function
 is about -0.2 and -0.3 respectively -- about ten per cent.
This justifies treating the gauge coupling as if it were not running and looking for
deviations.  For the more
familiar case of $SU(3)$ gauge group and $N_f=2$ fundamental fermions, the change
would be -0.8 and -2.0 at these couplings. (And in this case the observed coupling
runs faster than the perturbative result.\cite{DellaMorte:2004bc}.)

Looking ahead to my answer, I claim that $y_m\sim 1.5$ at the two values of
 the bare coupling where I measured it.

 I will investigate three different ways to measure $y_m$.The successful ones employ
finite size scaling arguments for the response of observables to the simulation volume.

The first way uses the correlation length directly.  When it grows to be the
 size of the simulation volume, Eq.~\ref{eq:corrlen} breaks down. Finite
 size scaling arguments allow us to use the breakdown to determine
the exponent. This test is done at two values of the bare coupling.

The two other tests involve working with extensive quantities. 
As $m_q$ is tuned to zero, the singular part of the free energy scales as
\bee
f_s(m_q) = m_q^{D/y_m}(A_1 + A_2  m_q^{|y_i|/y_m} +\dots).
\label{eq:fs}
\ee
where $D$ is the system's dimensionality (here $D=4$), and the prefactor
is just $1/\xi^D$ (the correlation length provides the appropriate dimensional factor).
 $A_1$ and $A_2$ are non-universal constants and $y_i$ is the 
biggest non-leading exponent.
This is most likely the exponent $y_g$ of the gauge coupling, $g^2-g^{2*}$, which
 can be determined from the beta function
as measured in (for example) Schrodinger functional simulations at $m_q=0$.
It is probably small; in Ref.~\cite{DeGrand:2009mt}, Hasenfratz and I estimated that
it was close to zero in this and related theories.

So the
 condensate, $\langle \bar \psi \psi \rangle =\Sigma(m_q)$,
scales with $m_q$ as
\bee
\Sigma(m_q) = \frac{\partial f_s}{\partial m_q} \sim m_q^{\alpha}
\ee
where
$\alpha = D/y_m -1$. This is exactly like the relation of the specific heat exponent to the
correlation length exponent at a conventional paramagnetic-ferromagnetic critical point.

Working with the condensate has the problem  that the particular lattice fermions used in
this simulation, Wilson-type fermions, explicitly break chiral symmetry 
in the action. This
introduces an additive shift to the condensate.
So I will use partial quenching: I will take configurations generated
with lattice fermions which do not have exact chiral symmetry, and measure the Dirac
spectrum using valence quarks which are
an implementation of lattice fermions with exact chiral symmetry (overlap fermions).
One can question whether the valence fermion sees a faithful realization of
what is happening in the equilibrium distribution of real dynamical variables.
In usual (low-$N_f$ fundamental QCD) this is believed to be the case. I think that
for a
first study, what I am going to do is adequate.
(It would of course be better to do simulations with chiral dynamical fermions,
as was done by the authors of Ref.~\cite{Fodor:2008hm}, but they are presently too expensive
for simulations at large volume.)

 The condensate has  UV-sensitive pieces
  \cite{Leutwyler:1992yt}, 
 $\langle \bar \psi \psi \rangle_{UV} \sim C_1 m_q + C_3 m_q^3 + \dots$
where $C_1 \sim 1/a^2$ and $C_3 \sim \log \  a$. This  masks the
$m_q^\alpha$ non-analytic behavior.
This is quite similar to the situation in finite temperature QCD, precisely at $T=T_c$,
where\cite{Karsch:2008ch}
\bee
\Sigma(a,m_q,T) \sim c_1 m_q/a^2 + c_\delta m_q^{1/\delta} + {\rm analytic}.
\label{eq:sqcd}
\ee
An IRFP theory is different from QCD in that  there are no Goldstone bosons
(which contribute their own non-analytic piece to the condensate, below $T_c$). It is also
different from QCD in that while in QCD, Eq.~\ref{eq:sqcd} applies only at $T_c$,
in an IRFP theory Eq.~\ref{eq:sqcd} gives the behavior of the condensate
throughout the basin of attraction of the IRFP.
Unfortunately, for my data set, the UV terms dominate $\Sigma(m)$.

A better way
to attack $y_m$ through the condensate involves the
Banks-Casher relation\cite{Banks:1979yr} between the condensate and the
 density of eigenvalues $\lambda$
 of the Dirac operator
$\rho(\lambda)$. At nonzero mass it is
\bee
\Sigma(m_q) = - \int \rho(\lambda) d \lambda \frac{2m_q}{\lambda^2+m_q^2}.
\ee
If the massless theory is conformal, and
the condensate $\Sigma(m_q) \sim m_q^{\alpha}$ for small mass, then $\rho(\lambda)$ also scales
as $\lambda^{\alpha}$.
A finite-size scaling argument\cite{Akemann:1997wi} 
 relates the scaling for the density $\rho$ to the scaling of
the value of individual eigenvalues. If we consider the average value of the $i$th eigenvalue
of the Dirac operator in a box of volume $V=L^D$, and  if  $\rho(\lambda)\sim \lambda^\alpha$,
then we expect
\bee
\langle \lambda_i \rangle \sim \left(\frac{1}{L}\right)^p
\label{eq:eigscale}
\ee
where the exponent is
\bee
p=\frac{D}{1+\alpha}.
\label{eq:scaling}
\ee
For the case of an IRFP theory, $p$ is equal to $y_m$, the leading
exponent.

(To derive this, note $\rho \sim \lambda^{\alpha}$ means that eigenvalues are uniformly distributed
in an $N=\alpha+1$ dimensional space of volume $V=R^N$,
\bee
\lambda= \frac{\pi}{R}(\sum_{i=1}^N n_i^2)^{1/2} \ \  n_i=1,2,\dots R 
\ee
 so  an eigenvalue scales as 
$\lambda_i \sim 1/R = (1/V)^\frac{1}{N}=(1/V)^\frac{1}{\alpha+1}$.
 Now suppose we are in $D$ physical dimensions;
in a box of volume $V$, there are  $V=L^D$ modes, from which Eq.~\ref{eq:scaling} is obtained.)

One example of this formula is free field theory: $\alpha=D-1$ and $p=1$. Another
is the case of chiral symmetry breaking encoded in the usual formulas of its Random Matrix Theory
analog: $\alpha=0$ so $\rho(\lambda) \rightarrow \rho_0$ a constant, and $p=D$. This is
$\langle \lambda_i\rangle \sim 1/V$, which is equivalent to the usual statement that the
 eigenvalue
spectrum depends on the dimensionless product $\lambda \Sigma V$.

For the case of a system which exhibits
chiral symmetry breaking, there is a tight theoretical description of the behavior of the lowest
eigenvalues of the Dirac operator, which allows one to relate delicate features of the
 spectrum to
the low energy constants of the theory (the condensate, the pseudoscalar decay constant, 
and possibly others).
This description is based on Random Matrix Theory (RMT).
However,  if a system is
 in a chirally-restored phase, there are no longer RMT predictions to compare results against.
Previous work shows  that our target theory has a weak-coupling phase
which is chirally-restored. The restoration of chiral symmetry is observed
through regularities in the spectrum of screening masses as well as the behavior of the
 pseudoscalar decay constant as a function of quark mass. Therefore, the
 analysis reported here only uses the simplest
 property of the eigenvalues, namely their scaling with system size.

 The paper proceeds as follows. In Sec.~\ref{sec:two} I describe details of the simulations.
In Sec.~\ref{sec:three}
I compute the correlation length exponent using finite size scaling.
Next, in Sec.~\ref{sec:four}
 I examine the mass dependence of the chiral condensate, and finally in
Sec.~\ref{sec:five}
I perform a scaling test for eigenvalues of the valence Dirac operator.
I conclude with a discussion of my results.

In an earlier preprint \cite{DeGrand:2009et}
 I tried to do several of the analyses I report on, in this paper.
In Ref.~\cite{DeGrand:2008kx} we observed that in the deconfined phase of sextet QCD, the
Polyakov loop ordered in one of the negative real directions,
 roughly along one of the complex elements of
$Z(3)$, ($\Re \langle  \Tr P(x)\rangle <0$, $\Im \langle  \Tr P(x) \rangle \ne 0$).
I believed that was the general situation for this theory and all the simulations were
done in those vacua.  After that paper appeared,
detailed studies of the phase structure (at smaller volumes) by
Machtey and Svetitsky \cite{MSS} showed that the true vacuum of the deconfined
phase is in fact the one in which the Polyakov loop is real and positive.
All the results of Ref.~\cite{DeGrand:2009et} only apply to metastable vacua.
Their conclusion renders it too uninteresting to publish.

\section{Numerical techniques and background\label{sec:two}}
I  performed simulations on a system with $SU(3)$ gauge fields and two flavors of 
dynamical fermions
in the symmetric (sextet) representation of the color gauge group.
The lattice action is defined by the single-plaquette gauge action and a
Wilson fermion action with added clover term~\cite{Sheikholeslami:1985ij}.
The fermion action employs
the differentiable hypercubic smeared link of
 Ref.~\cite{Hasenfratz:2007rf}, from which the symmetric-representation gauge connection
for the fermion operator is constructed. No tadpole-improvement is used and the
clover coefficient is set to its tree-level value.
The smearing parameters for the links are the same as in
Ref.~\cite{Hasenfratz:2007rf}, $\alpha_1=0.75$, $\alpha_2=0.6$, $\alpha_3=0.3$.
The bare parameters which are inputs to the simulation are the gauge 
coupling $\beta=6/g^2$  and the fermion hopping parameter $\kappa$.
 The integration is done with  one additional heavy
pseudo-fermion field as suggested by Hasenbusch \cite{Hasenbusch:2001ne},
multiple time scales \cite{Urbach:2005ji},
and a second-order Omelyan integrator \cite{Takaishi:2005tz}.

The routines for simulating sextet-representation fermions were developed
with (and mostly by) B.~Svetitsky and Y.~Shamir. The dynamical fermion algorithm was adapted
from a program written by A.~Hasenfratz, R.~Hoffmann and S.~Schaefer\cite{Hasenfratz:2008ce}
All computer code is based on the publicly available package of the MILC collaboration~\cite{MILC}.

Simulation volumes range up to $16^4$ sites, 
and typical data sets range from
a few hundred to a thousand trajectories. I recorded lattices every five trajectories 
(of unit simulation time trajectory length) and collected 40 lattices per parameter set
for the calculation of spectral observables  and  overlap eigenvalues.

Correlation lengths are taken to be inverses of 
 screening masses, from
correlators of operators at different separations  measured along one of the spatial
directions of the lattice.
The trick of combining periodic and anti-periodic boundary conditions for valence quarks
\cite{Blum:2001xb,Aoki:2005ga,Allton:2007hx,Aubin:2007pt}
is used in  these measurements.

Throughout this work, instead of quoting $\kappa$, I will use the the Axial Ward Identity (AWI) quark mass,
 defined through
\bee
\partial_t \sum_x \svev{A_0(x,t)X(0)} = 2m_q \sum_x \svev{ P(x,t)X(0)}.
\label{eq:AWI}
\ee
where $A_0=\bar \psi \gamma_0\gamma_5 \psi$, $P = \bar \psi \gamma_5 \psi$, and $X$ is any source.
The derivative is taken to be the naive  difference
operator ($\partial_\mu f(x)=(f(x+\hat\mu a) - f(x-\hat\mu a))/(2a)$).
For $X$ I used a Coulomb gauge-fixed Gaussian source.

Machtey and Svetitsky \cite{MSS} have performed careful studies of
the phase structure of this theory and have shown that that the true vacuum (in the deconfined 
phase) is the
one in which the expectation value of the Polyakov loop is real and positive.
All simulations reported here were done in this vacuum.

The valence Dirac operator whose eigenvalues are used in Sec.~\ref{sec:five}
 is the overlap
operator \cite{Neuberger:1997fp,Neuberger:1998my}.
 Details of the particular implementation of the action are described in
 Refs.~\cite{DeGrand:2000tf,DeGrand:2004nq,DeGrand:2006ws,DeGrand:2007tm,DeGrand:2006nv}.
The only new ingredient is the application to symmetric-representation fermions, using the
same combination of hypercubic link and projection into the fermionic representation
 as for the dynamical fermions.
Eigenvalues of the squared Hermitian Dirac operator $D^\dagger D$ are computed
 using the ``Primme'' package of
McCombs and Stathopoulos\cite{primme} and split apart in the usual way.
Eigenvalues are quoted after stereographic projection.

There is a potential problem with this analysis involving the
index theorem, relating the winding number of the gauge field $Q$
to the number of Dirac fermion zero modes,
\bee
{\rm index} = 2T(R) Q.
\ee
 $T(R)$ is the Dynkin index of the representation $R$, 1/2 for fundamental representation fermions,
 $(N+2)/2$ for two-index symmetric representation
 fermions in the color group $SU(N)$, and so on.
 Thus we expect to see
multiples of 5  zero modes for sextet overlap fermions in our $SU(3)$ case.

However, ten years ago Heller, Edwards, and Narayanan discovered\cite{Heller:1998cf}
 that the index theorem applied
to adjoint overlap fermions in background $SU(2)$ gauge configurations failed: while $2T(R)=4$,
 they saw
configurations with zero modes which were not multiples of four. More recently, 
similar results were reported by
Garcia Perez, Gonzalez-Arroyo and Sastre\cite{GarciaPerez:2007ne}.
 In simulations where the bare gauge coupling is large, I have seen
configurations whose zero mode content was a not a multiple of 5.

Fortunately, the authors of Ref.~\cite{Fodor:2009nh} have studied the index theorem for
 sextet representation fermions in quenched background $SU(3)$ gauge field configurations 
and shown that
in the continuum limit only configurations with multiples of five zero modes are found.
The ``fractional states'' are apparently  just a particular failure of the overlap action to capture
topology when the gauge configuration is rough.

This is not a problem for this project. The
 gauge configurations at the parameter value used to compute eigenvalues
($\beta=5.2$)  were smooth enough that
all  of the lattices I collected  had $Q=0$.
I performed some trial simulations at a lower $\beta$ of 4.8 and also saw only $Q=0$.
However, for these rougher configurations the cost of the overlap operator went up by
about a factor of four, and it did not seem like a good use of computer time to continue
running there.

\begin{figure}
\begin{center}
\includegraphics[width=0.6\textwidth,clip]{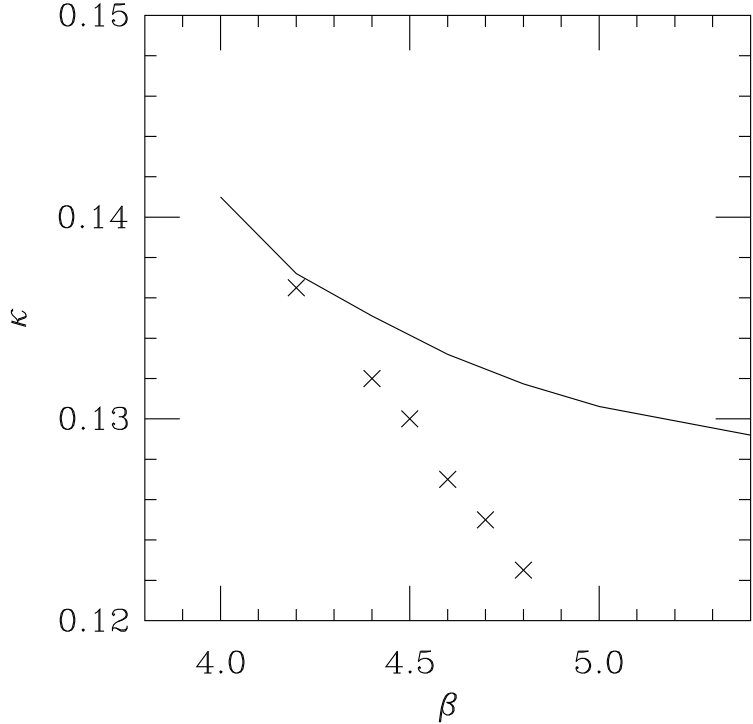}
\end{center}
\caption{Map of the bare coupling constant plane relevant to our sextet simulations. The
solid line
is the line of zero quark mass, $\kappa=\kappa_c$. The crosses show the location of the
 confinement-deconfinement crossover at $N_t=6$.
\label{fig:betakappa}}
\end{figure}

A map of the simulation region is shown in Fig.~\ref{fig:betakappa}.
It is qualitatively similar to what we found with another bare action
\cite{DeGrand:2008kx}.
The line is the location in bare parameter space where the AWI quark mass vanishes,
$\kappa_c$. The crosses show the location of the $N_t=6$ deconfinement phase transition.
To the left of this line, the system is confined and seems to be chirally broken.
However, this region is little explored. It is unknown how (or if) the deconfinement
line attaches to the $\kappa_c$ line.
To the right of the line, the system is deconfined on all observed volumes
and spectroscopy shows parity doubling. This plus the smallness of the
pseudoscalar decay constant leads me to conclude that the system is in a chirally restored
state throughout that phase. All data used in this study comes from the weak coupling phase.

Table \ref{tab:Table1} lists all simulation points use in this study. Since I will
report results in terms of the AWI quark mass, I include it in the table. When there are
several volumes, the quoted mass is from the smallest volume. However, generally the volume
dependence
of the quark mass is small ( a few digits in the least significant figure).
In the analysis, data from any particular volume is plotted and used at its measured quark mass
in that volume.  As $\beta$ falls, the available parameter
space in the deconfined phase shrinks to a smaller and smaller range of quark masses.

\begin{table}[t]
\caption{ Simulation points reported in this work. Since I am using the AWI quark mass as
my independent
variable, I catalog it along with the bare parameters ($\beta$, $\kappa$, volume).
\label{tab:Table1}}
\begin{ruledtabular}
\begin{tabular}{cccc}
$\beta$ & $\ \kappa\ $ & $\ am_q\ $ & volumes \\
\hline
4.20 & 0.1370 & 0.110 & $12^3\times 6$   \\
\hline
4.40 & 0.1335 & 0.106 & $12^3\times 6$   \\
4.40 & 0.1345 & 0.052 & $12^3\times 6$   \\
\hline
4.60 & 0.1300 & 0.165 & $12^3\times 6$,  $12^4$   \\
4.60 & 0.1320 & 0.065 & $12^3\times 6$,  $12^4$   \\
4.60 & 0.1325 & 0.027 & $12^4$   \\
\hline
4.80 & 0.1230 & 0.412 & $12^3\times 6$, $12^3\times 8$,  $12^4$   \\
4.80 & 0.1250 & 0.300 & $12^3\times 6$, $12^3\times 8$,  $12^4$   \\
4.80 & 0.1260 & 0.233 & $12^4$   \\
4.80 & 0.1270 & 0.209 & $12^3\times 6$, $12^3\times 8$,  $12^4$   \\
4.80 & 0.1285 & 0.146 & $12^3\times 6$, $12^4$, $16^4$   \\
4.80 & 0.1290 & 0.104 & $12^3\times 8$, $12^4$,  $16^4$   \\
4.80 & 0.1300 & 0.086 & $12^3\times 6$, $12^3\times 8$,  $12^4$, $16^4$   \\
\hline
5.20 & 0.1100 & 0.940 & $12^3\times 6$,  $12^4$   \\
5.20 & 0.1150 & 0.630 & $12^3\times 6$,  $12^4$   \\
5.20 & 0.1220 & 0.310 & $12^3\times 6$,  $12^4$   \\
5.20 & 0.1235 & 0.220 & $12^4$   \\
5.20 & 0.1250 & 0.195 & $12^3\times 6$,  $12^4$, $16^4$   \\
5.20 & 0.1270 & 0.096 & $12^4$   \\
5.20 & 0.1285 & 0.066  &$ 8^4$, $10^4$,  $12^3\times 6$, $12^3\times 8$,  $12^4$,
$16^3\times 8$,
$16^4$   \\
5.20 & 0.1290 & 0.047 & $12^3\times 6$,  $12^4$, $16^4$   \\
\end{tabular}
\end{ruledtabular}
\end{table}

\section{Finite-size scaling\label{sec:three}}
\begin{figure}
\begin{center}
\includegraphics[width=0.8\textwidth,clip]{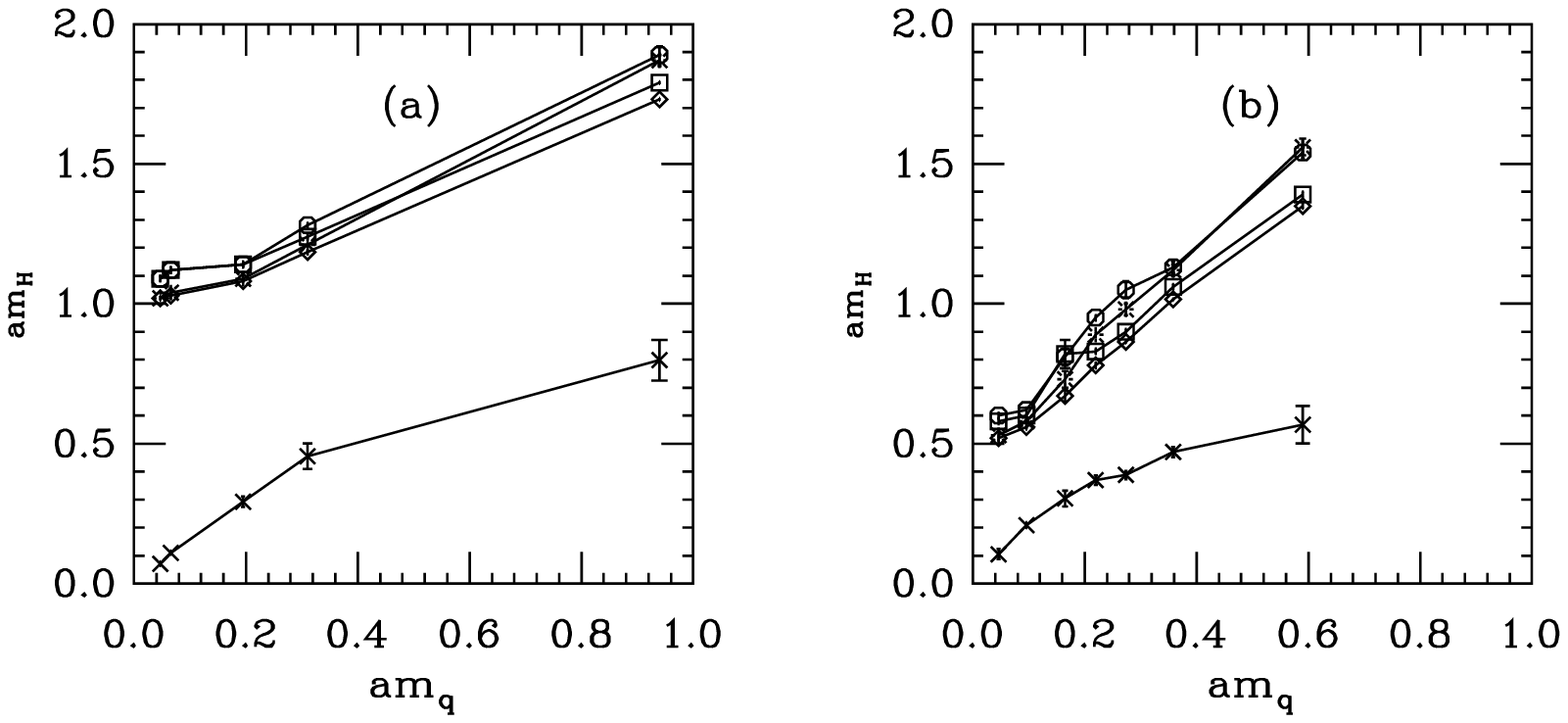}
\end{center}
\caption{
Screening masses at $\beta=5.2$ for two simulation volumes.
Left: $12^3\times 6$, right, $12^4$. Symbols show mass vs AWI quark mass:
diamonds -- pseudoscalar; squares -- vector; octagons -- axial vector; burst -- scalar; 
crosses --  pseudoscalar decay constant $f_{PS}$.
\label{fig:m5.2}}
\end{figure}

I begin this section with several pictures of spectroscopy to set the stage.
Fig.~\ref{fig:m5.2} are plots of screening masses at $\beta=5.2$ on $12^3\times 6$ and
$12^4$ volumes. These two panels show  that in the weak coupling
phase the excitation spectrum is parity doubled and that the pseudoscalar decay constant
becomes small as the quark mass vanishes.

\begin{figure}
\begin{center}
\includegraphics[width=0.8\textwidth,clip]{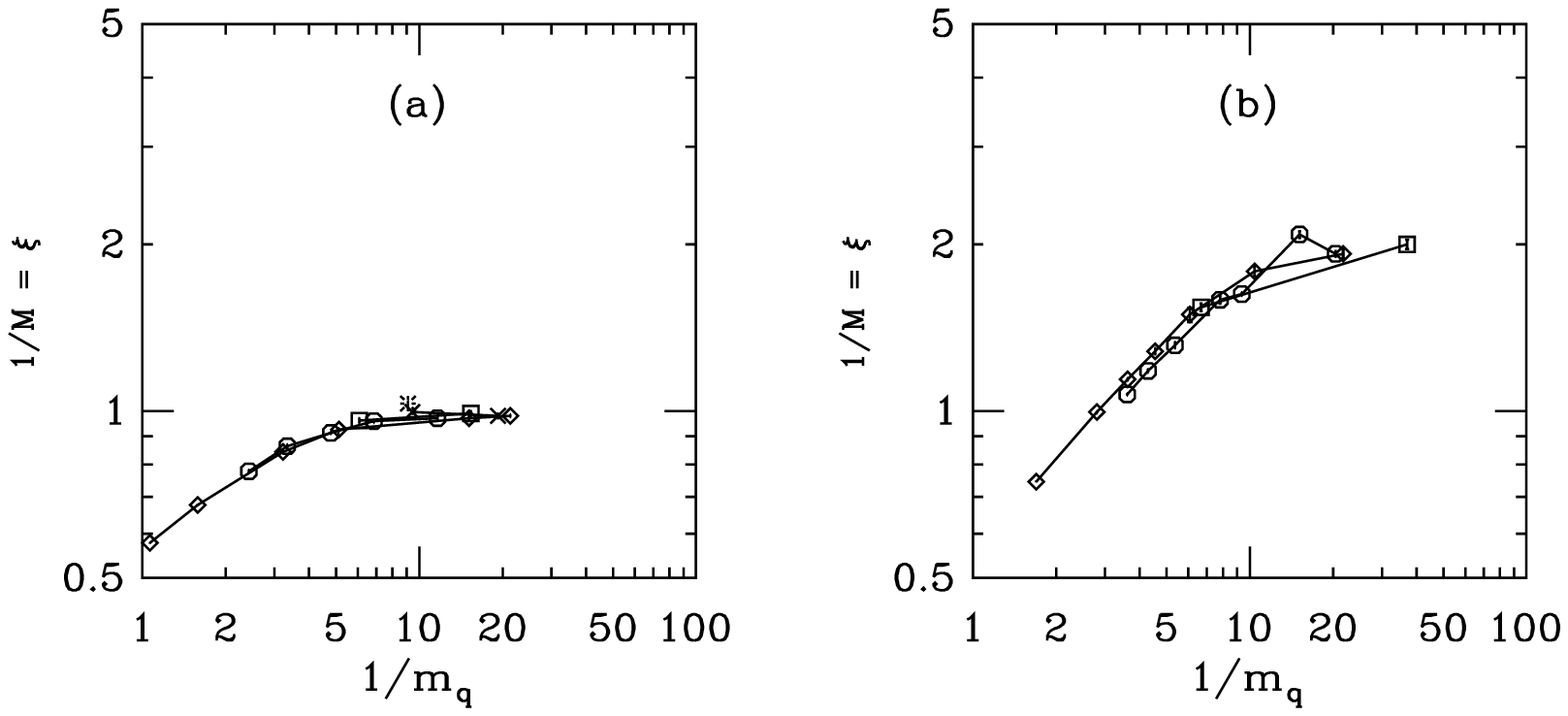}
\end{center}
\caption{ Correlation length (inverse pseudoscalar screening mass) versus inverse AWI
quark mass, on $12^3\times 6$ (left) and $12^4$ volumes (right). Plotting symbols are
bursts, $\beta=4.2$;
crosses, $\beta=4.4$;
squares, $\beta=4.6$;
octagons, $\beta=4.8$;
diamonds, $\beta=5.2$.
\label{fig:xivsm612}}
\end{figure}

Throughout the weak coupling phase, masses depend strongly on the quark mass
and weakly on the bare gauge coupling. Fig.~\ref{fig:xivsm612} illustrates this feature
for the correlation length, defined as the inverse of the pseudoscalar screening mass,
 showing $12^3\times 6$ volumes on the left and $12^4$ volumes
on the right. Different plotting symbols correspond to different $\beta$ values.

\begin{figure}
\begin{center}
\includegraphics[width=0.6\textwidth,clip]{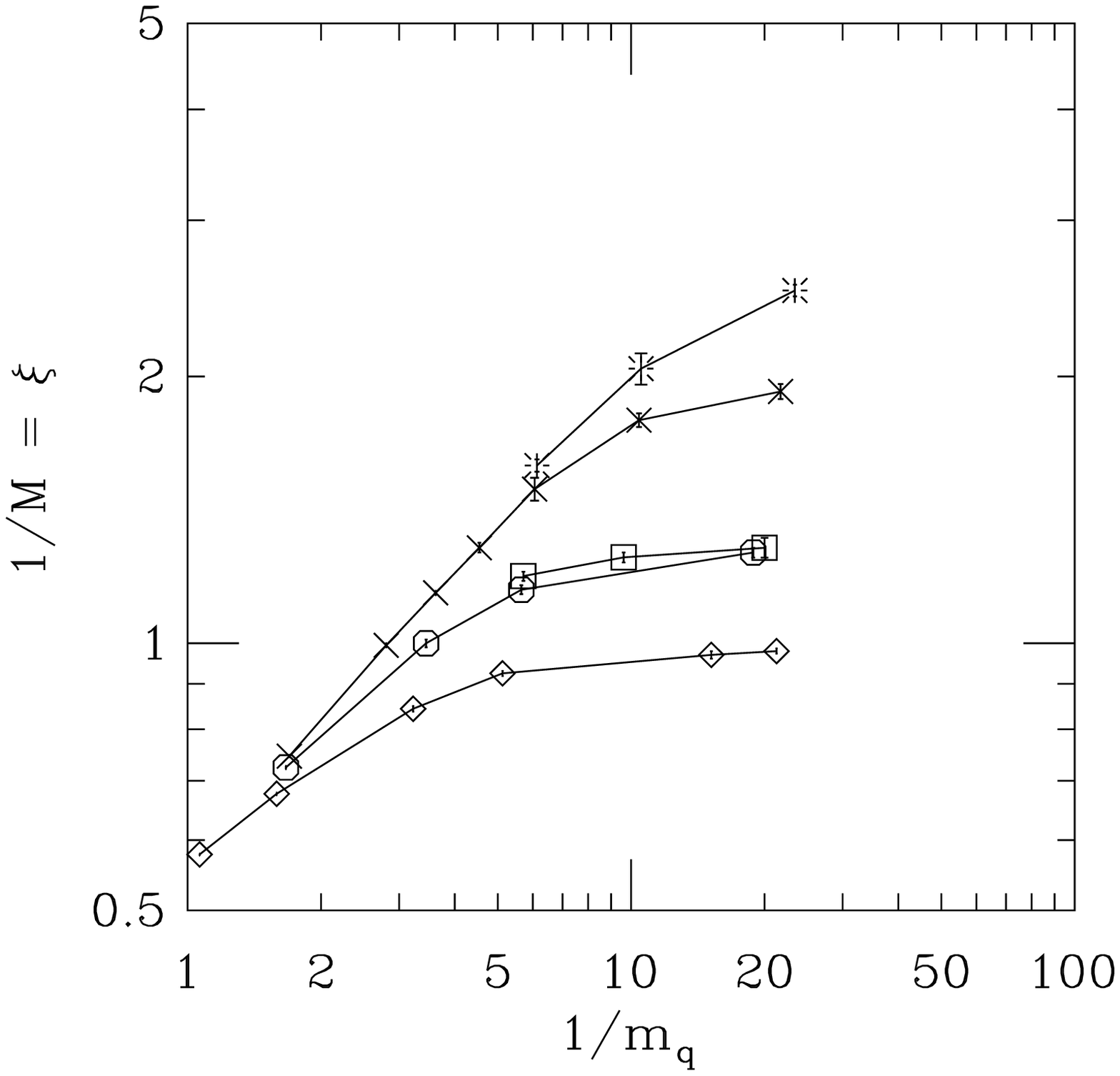}
\end{center}
\caption{Correlation length (inverse pseudoscalar screening mass) versus inverse
AWI quark mass at $\beta=5.2$. Plotting symbols are for different simulation volumes,
diamonds, $12^3\times 6$;
octagons, $12^3\times 8$;
squares, $16^3\times 8$;
crosses, $12^4$;
bursts, $16^4$.
\label{fig:xivsm}}
\end{figure}

Finally, I combine data from many volumes at one gauge coupling, $\beta=5.2$,
in Fig.~\ref{fig:xivsm}. The saturation of the correlation length at a scale
proportional to the temporal length of the lattice is apparent. Superficially, this is
just the Matsubara cutoff $M \sim 2\pi/N_t$ expected from the fermions' antiperiodic temporal
 boundary conditions.

The relation between correlation length $\xi$ and quark mass given by Eq.~\ref{eq:corrlen}
is only expected to hold when the system size $L$ is much larger than $\xi$.
When the correlation length measured in a system of size $L$
(call it $\xi_L$) becomes comparable to $L$, $\xi_L$ saturates at $L$ even as $m_q$
vanishes.
However, if the only large length scales in the problem are $\xi$ and $L$,
then overall factors of length can only involve $\xi$ and $L$. For the correlation length itself,
this argument says that
\bee
\xi_L = L F(\xi/L)
\ee
where $F(x)$ is some unknown function of $\xi/L$. A somewhat more useful version of this
relation can be written by using Eq.~\ref{eq:corrlen} to say
\bee
\xi_L = L f(L^{y_m} m_q)   .
\label{eq:fss2}
\ee
Then one can plot $\xi_L/L$ vs $L^{y_m} m_q$ for many $L$'s, vary $y_m$, and look for the
appearance of a smooth curve. The data from different $L$'s will march across the x axis
at different rates.

\begin{figure}
\begin{center}
\includegraphics[width=0.8\textwidth,clip]{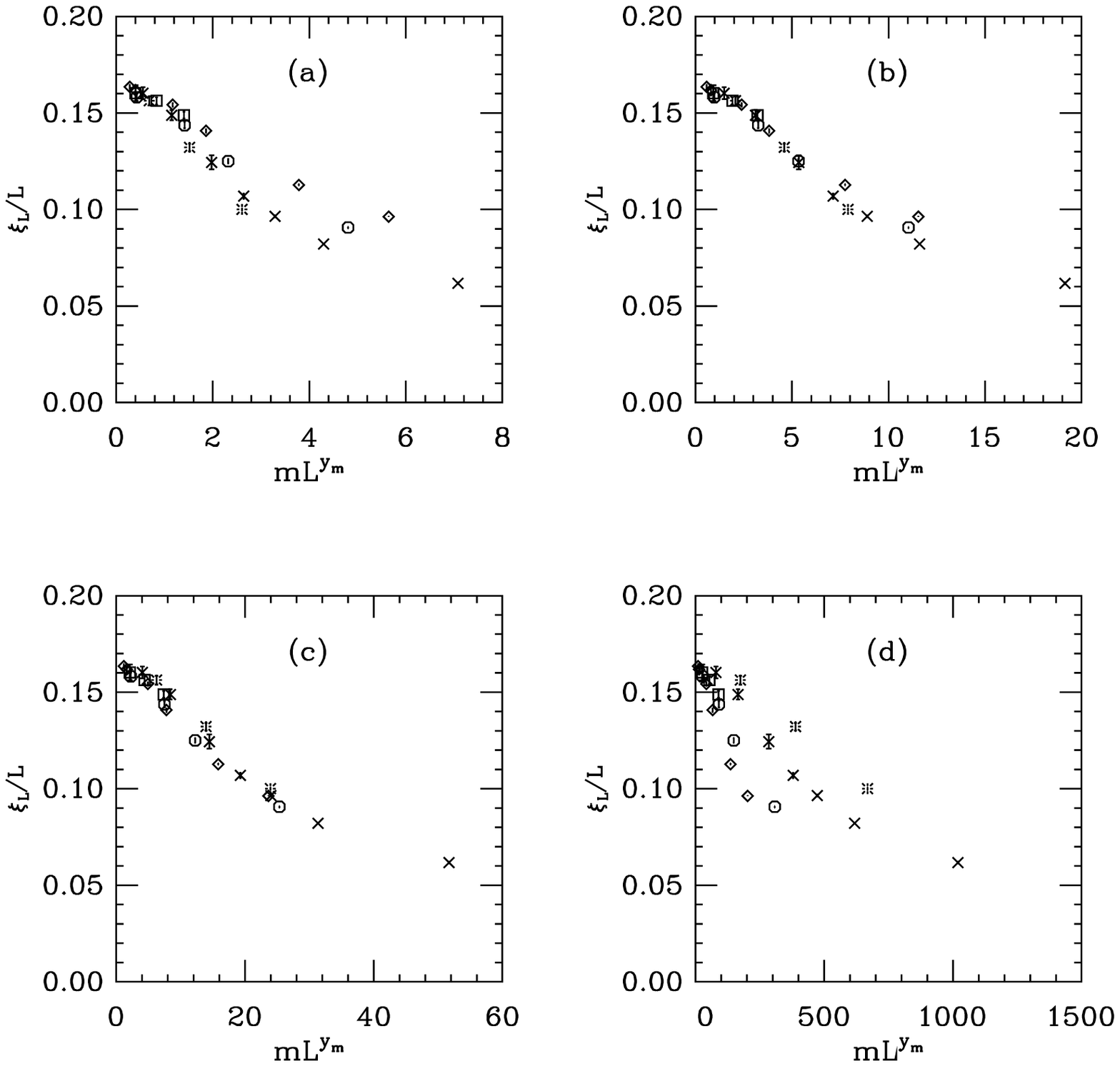}
\end{center}
\caption{
Plots of $\xi_L/L$ vs $m_q L^{y_m}$ at $\beta=5.2$ for four choices of $y_m$:
(a) $y_m=1.0$, (b) $y_m=1.4$, (c) $y_m=1.8$ (d) $y_m=3.0$.
Plotting symbols are for different simulation volumes,
diamonds, $12^3\times 6$ ($L=6$);
octagons, $12^3\times 8$ ($L=8$);
squares, $16^3\times 8$ ($L=8$);
crosses, $12^4$ ($L=12$);
bursts, $16^4$ ($L=16$).
\label{fig:alltest}}
\end{figure}

A good data set to use is the one of Fig.~\ref{fig:xivsm},
$\beta=5.2$. These are screening masses, so I will take
$L=N_t$ regardless of the value of $N_s$. A scan of Eq.~\ref{eq:fss2}
 is shown in Fig.~\ref{fig:alltest}.
 Already one can see that a choice of $y_m\sim 1.5$ pulls all the data from
Fig.~\ref{fig:xivsm}
onto a single curve.

To turn this observation into a number with an uncertainty is a bit awkward.
We are not doing a fit, because the function $f(x)$ of Eq.~\ref{eq:fss2} is unknown.
Instead I will do the analysis in two different ways:

First \cite{BQSidea} 
take the full range of values of $\xi_L/L$ and slice it into a set of $N_b$ bins.
In the $j$th bin, define
\bee
\langle x_j \rangle = \frac{1}{N_i}\sum_{i\in bin} m_i L_i^{y_m}
\ee
and
\bee
\langle x_j^2 \rangle = \frac{1}{N_i}\sum_{i\in bin} (m_i L_i^{y_m})^2.
\ee
Then define a goodness-of-fit parameter as
\bee
\chi^2 = \sum_{j=1}^{N_b} \left( \frac{\langle x_j^2\rangle}{\langle x_j \rangle^2} -1
\right).
\ee
Minimize this function with respect to $y_m$ and fold the whole procedure into
a jackknife over all the different mass-$L$ combinations to get an uncertainty.
The division by $\langle x_j \rangle^2$ mimics the eye's attempt to minimize the
fractional spread of the points about their average.
Notice that bins with only a single entry do not contribute to $\chi^2$.
Binning the data introduces the possibility of a dependency on the number of bins.
I assign an error by jackknifing the data set.
In practice, if there are too many bins, there are seldom bins with more than one point,
and the jackknife error becomes large.

A second approach follows the method of Ref.~\cite{BhS}. The idea is to use each data set
(each different $L$ value) to estimate $f(x)$ and to find the $y_m$ which pulls the
other $L$ sets onto it. This is done inclusively; all data sets
take a turn at being the fiducial. The quantity to minimize is
\bee
P(y_m) = \frac{1}{N_{over}} \sum_p \sum_{j\ne p} \sum_{i,over}
\left( \frac{\xi_L(m_{i,j})}{L_j} - f_p(L_p^{y_m}m_{ij})\right)^2
\ee

The interpretation of this long formula is that data set $p$ is used to
estimate the scaling function $f(x)$. This is done by interpolation, either by polynomials or
rational functions, using the recorded values of $\xi_L/L$. The label ``over'' indicates
that the sum only includes data from set $j$ whose $x$ values, $L_j^{y_m}m_{i,j}$,
overlap the range of $x$'s of set $p$.  The overall
factor
of $1/N_{over}$ counts the total number of points and guards against recording a zero value
of $P$ if there are no overlap.   (Bhattacharjee and Seno actually consider powers
other than two inside the sum.) $P$ is minimized by the optimal $y_m$.
 This method has an advantage over the first one that
one does not need to bin the data, and a disadvantage that the interpolation algorithm has
to be robust. This can be a problem for extreme values of $y_m$. The number $N_{over}$
varies as $y_m$ is tuned. This could potentially make  $P$ discontinuous. 

The two methods produce similar results. At $\beta=5.2$ there are 23 separate data points
(values of $m_q$ and $L$), with $\xi_L/L$ ranging from about 0.08 to 0.16.
 Asking for 5 bins gives (on average) 4 useful bins and $y_m=1.43(25)$. Asking for 6 bins
gives 5 useful bins and $y_m=1.44(24)$. Asking for 8 bins produces on average 5 useful bins
and $y_m=1.51(38)$. With the second method, $y_m=1.53(13)$ from a single-elimination jackknife
over the data set.

Bhattacharjee and Seno advocate taling an error from an approximation to the second
derivative of P,
\bee
\Delta y_m = \eta y_m\left( 2 \ln \frac{P(y_m(1+\eta))}{P(y_m)}\right)^{-1/2}
\ee
With $\eta=0.1$ this produces $y_m=1.54(11)$.

\begin{figure}
\begin{center}
\includegraphics[width=0.8\textwidth,clip]{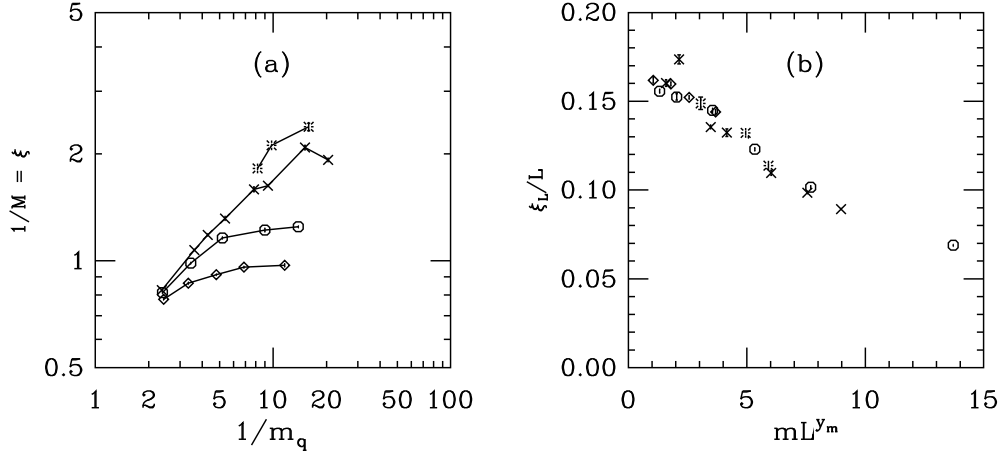}
\end{center}
\caption{
Plot of (a) $\xi_L$ vs $L$ and (b)
 $\xi_L/L$ vs $m_q L^{y_m}$ at $\beta=4.8$ for  $y_m=1.4$:
Plotting symbols are for different simulation volumes,
diamonds, $12^3\times 6$ ($L=6$);
octagons, $12^3\times 8$ ($L=8$);
crosses, $12^4$ ($L=12$);
bursts, $16^4$ ($L=16$).
\label{fig:fss4.8}}
\end{figure}

At $\beta=4.8$ the situation is similar, but noisier. I again have four $L$'s, from
$12^3\time 6$, $12^3\times 8$, $12^4$ and $16^4$ volumes ($L=6$, 8, 12, 16 respectively) and a
total
of 21 mass  and size combinations. The data is noisier than the $\beta=5.2$ data set.
Fig. \ref{fig:fss4.8} shows the data before and after collapse to a line.
A five-bin single jackknife fit gives $y_m=1.21(32)$ while the second method gives
1.41(26).   The approximate second derivative gives $y_m=1.40(20)$.
A conservative summary of  values and uncertainties is
 $y_m=1.5(2)$ at $\beta=5.2$ and 1.4(2) at $\beta=4.8$.

In this analysis, I made a particular choice of a geometry and a method of defining a correlation
function
(screening masses, sensitive to the length of the antiperiodic boundary condition in the temporal
direction). I do not believe that it is better than alternative methods one could try (for
example, taking $L^3\times L_t$ volumes and varying $L$). People interested in finite size
scaling tests should try other geometries.  

In the statistical mechanics finite size scaling literature it is common to
perform finite size scaling analyses on susceptibilities. I have experimented
with this, but probably not extensively enough. Mocking up susceptibilities by integrating
over correlation functions ($\chi \sim \sum_x \langle O(x)O(0)\rangle$) has not produced
interesting peaks, because the correlators -- and their integrals -- are dominated
by short distance effects. This deserves more study.

\section{The condensate directly from the overlap operator\label{sec:four}}
In this section and the next one, I only have data at $\beta=5.2$.
I ran at one $\kappa$ value, $\kappa=0.1285$. This corresponds to a small AWI mass.
 As I remarked, it
was quite expensive to evaluate the overlap operator at stronger coupling.

\begin{figure}
\begin{center}
\includegraphics[width=0.8\textwidth,clip]{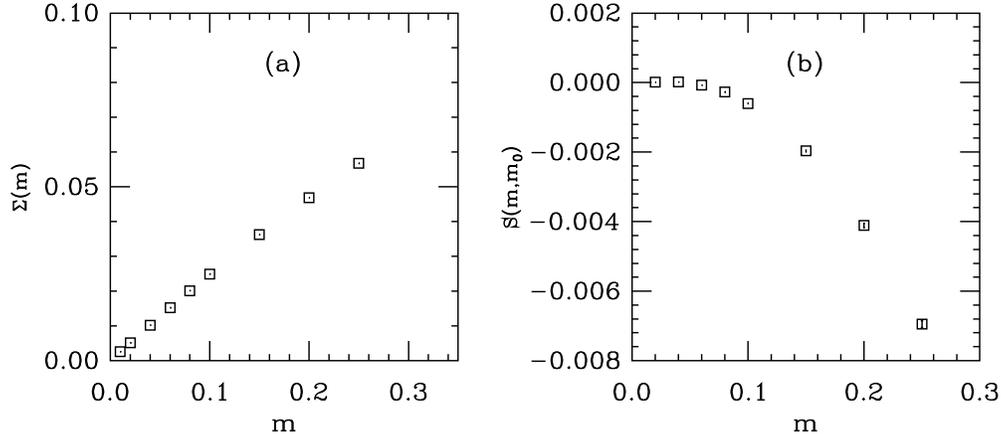}
\end{center}
\caption{(a) Condensate and (b) $S(m,m_0)$ from valence overlap fermions
as a function of the valence overlap mass, from background configurations at $\beta=5.2$,
$\kappa=0.1285$, on $12^4$ volumes.
\label{fig:condensate}}
\end{figure}

The condensate is measured  in the usual way, with a vector of Gaussian random numbers
defined on every site of the lattice, $\eta_i$, 
through an average over the random vectors,
\bee
 \Sigma(m_q) =\frac{1}{N}\sum_i \Sigma_i = \frac{1}{N}\sum_i \eta^\dagger_i \hat
D(m_q)^{-1} \eta_i
\ee
and as usual for the overlap operator, $\hat D(m_q)^{-1}$ is the subtracted, shifted
Dirac operator $\hat D(m_q)^{-1}= (D(m_q)^{-1} - 1/(2r_0))/(1-m_q/(2r_0)) $.
I accelerated the inversion by deflating with the eight smallest eigenmodes of $D^\dagger
D$. This is all quite standard \cite{DeGrand:2000tf}.
Even a small data set (twenty lattices, twelve random vectors per lattice)
produces a nice signal (Fig.~\ref{fig:condensate}a), which is 
almost completely linear in the quark mass. This demonstrates rather dramatically that the gauge
field configurations at one point in the weak coupling phase do not allow
chiral valence quarks to form a condensate. The linear behavior
 is just the UV-dominated
(proportional to $m_q/a^2$) term in the expected expansion.
 I tried to remove it using a variation on a trick from
Ref.~\cite{Bazavov:2009zn}:
Compute $\Sigma(m_q)$ for each random number, for many masses simultaneously
by using a multi-mass
sparse matrix inversion algorithm. Then pick a mass as a fiducial and compute
the difference $S(m_q,m_0)_i=\Sigma(m_q)_i-(m_q/m_0)\Sigma(m_0)_i$
 random number by random number.
Finally, form the average over random seeds. The high correlations in the values of the
condensate for different masses are removed from the error budget in $S(m_q,m_0)$.
I took nine masses ranging from $am_q=0.01$ to 0.25 and used $am_0=0.01$.
The result is shown in Fig.~\ref{fig:condensate}b.

Unfortunately, a fit of $S(m_q,m_0)$ to a power, $m_q^\alpha$, produces a noisy
result that $\alpha \sim 3$. The exponent is unstable against the set of masses 
included in the fit. This is most likely (and one could not argue that it is not)
 just the nonleading UV term
($m_q^3\log a$). Thus this attempt fails.
Something else is required, which is insensitive to the UV part of the condensate.
That follows in the next Section.

\section{Finite size scaling of Dirac eigenvalues\label{sec:five}}
\begin{figure}
\begin{center}
\includegraphics[width=0.7\textwidth,clip]{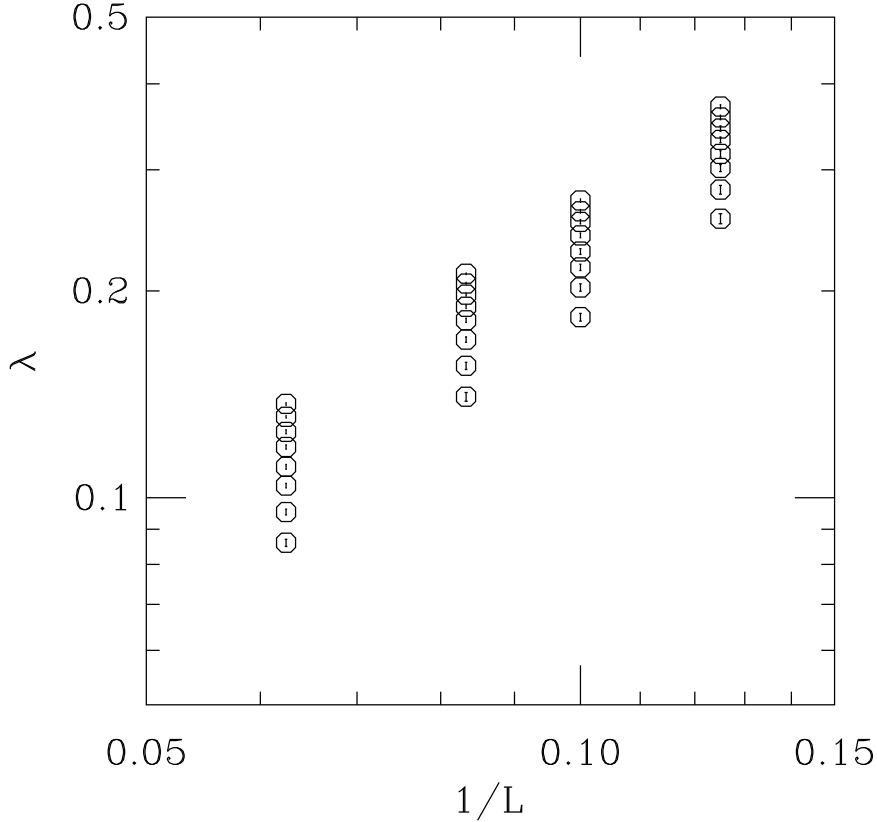}
\end{center}
\caption{Average values of 8 lowest eigenvalues of the sextet-representation
valence overlap operator, vs $1/L$,
where the lattice volume $V$ is defined to be $V=L^4$. The actual volumes, moving
from left to right across the graph, are $16^4$,
$12^4$, $10^4$ and $8^4$.
}
\label{fig:eigvsl}
\end{figure}

To test the scaling law of Eq.~\ref{eq:eigscale} I generated ensembles of lattices
at one set of bare parameters, $\beta=5.2$ and $\kappa=0.1285$, and a number of
simulation volumes, and computed the lowest eight eigenvalues of the massless overlap
Dirac operator on them.
The bare parameter set was chosen to have a light valence quark mass
and to be within the weak coupling phase.
All of the configurations collected at this parameter value have zero topological charge.
To see scaling, it is necessary to preserve the geometry
of the simulation volume (to compare, for example $L^4$ lattices at different $L$'s).
My primary data set is $L^4$ lattices with $L=8$, 10, 12, and 16.
The resulting eigenvalue spectrum is shown in Fig.~\ref{fig:eigvsl}. By eye, it 
seems to show power law behavior.

\begin{figure}
\begin{center}
\includegraphics[width=0.8\textwidth,clip]{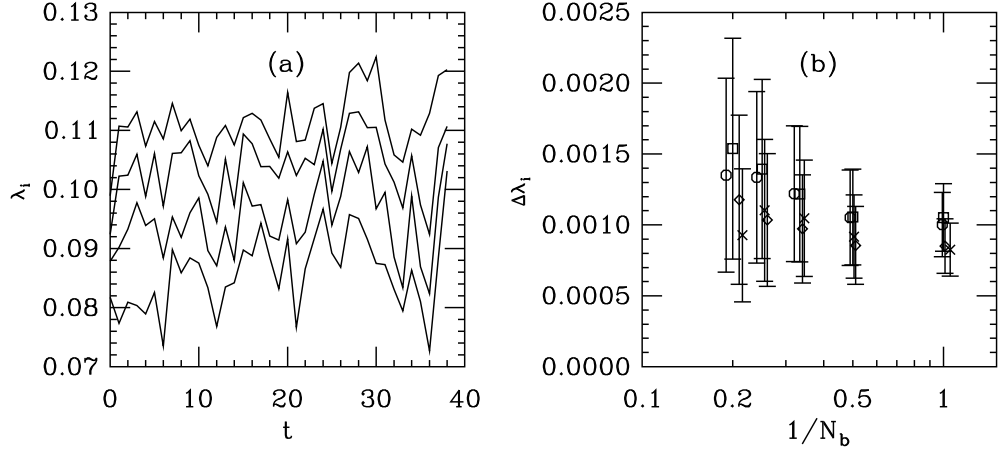}
\end{center}
\caption{(a) Time history of overlap eigenvalues from the $16^4$ data set (in units of 5
HMC time steps).
(b) Uncertainty on the average $\langle \lambda_i \rangle$ as a function of bin size.
Symbols are squares for $i=1$,
octagons for $i=2$,
diamonds for $i=3$,
and crosses for $i=4$.}
\label{fig:example}
\end{figure}

Fig~\ref{fig:example} illustrates the quality of the data.  The left panel shows
simulation time histories of the lowest four eigenvalues of the $16^4$ data set.
Each measurement is separated by five HMC trajectories. The right panel
shows the error on the average computed by blocking $N_b$ successive measurements
together.
The lowest eigenvalue is clearly the noisiest, but the autocorrelation time 
does not seem  to be too large:
I bin two successive lattices ($N_b=2$ or $\Delta t = 10$ HMC units) together before
averaging.

Now I wish  to extract an exponent from Fig. \ref{fig:eigvsl}.  For
theoretical input, all we have is Eq.~\ref{eq:eigscale}, the finite-size scaling
formula.
One does not know a priori if it applies to all the eigenvalues, only to the lowest
eigenvalues,
or if there is some minimum volume for which it applies.
I will just proceed empirically: I will look at fits to individual eigenvalues, then
groups of them.
I will fit all the data sets or drop the smallest volume and fit only the larger ones.

\begin{figure}
\begin{center}
\includegraphics[width=0.7\textwidth,clip]{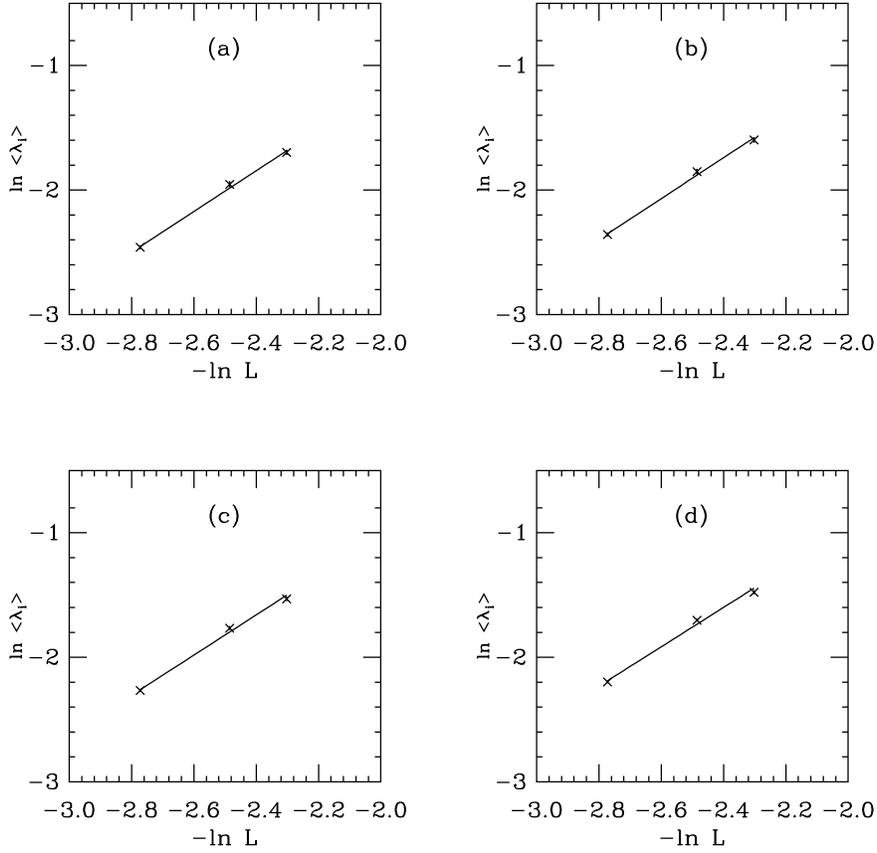}
\end{center}
\caption{Three-volume fits to individual eigenvalues, (a) the lowest eigenvalue
(b) the first excited state (c) the second excited state (d) the third excited state.
The volumes are (from the left) $16^4$, $12^4$, and $10^4$.
}
\label{fig:several3}
\end{figure}

\begin{figure}
\begin{center}
\includegraphics[width=0.7\textwidth,clip]{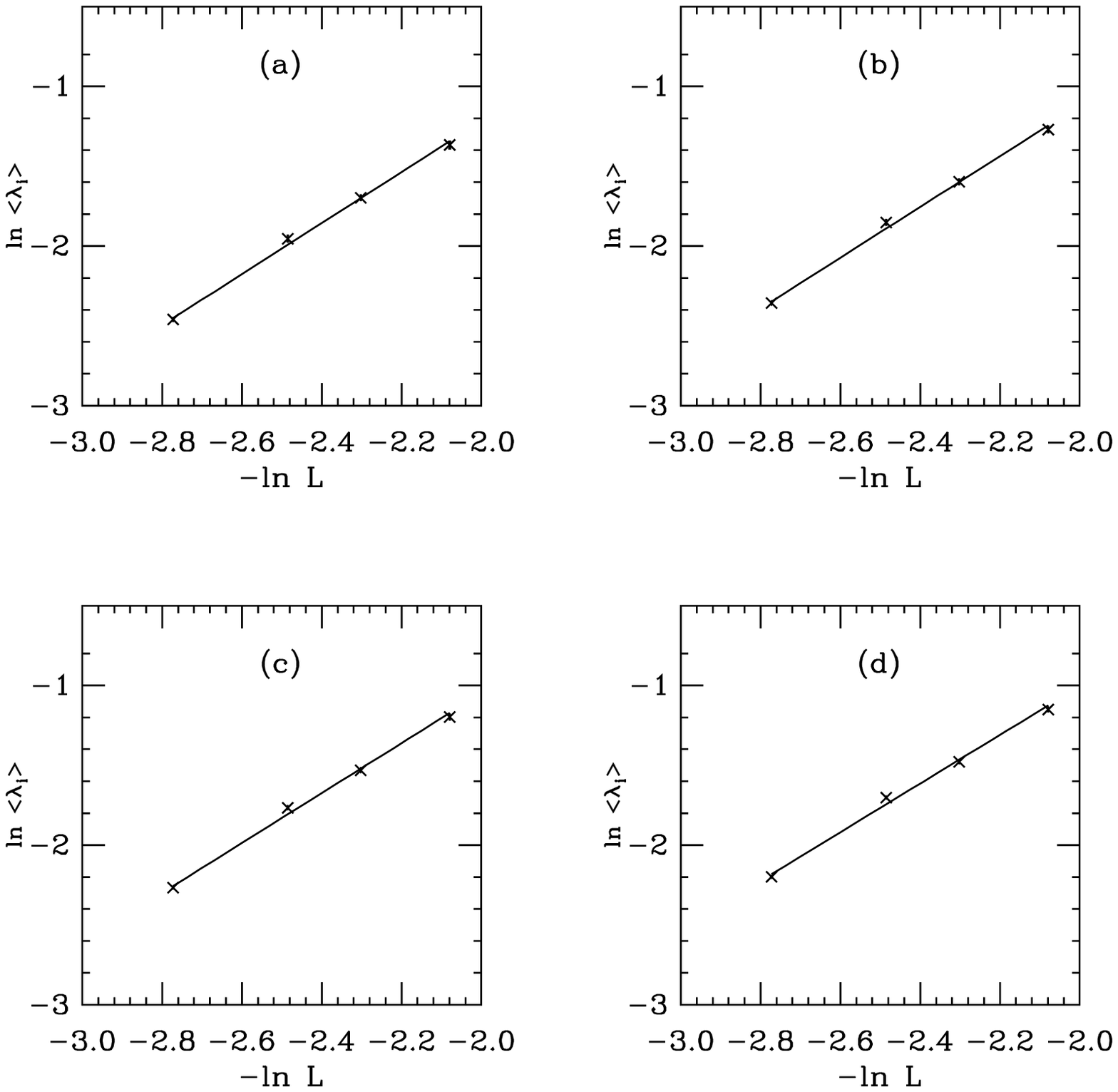}
\end{center}
\caption{Four-volume fits to individual eigenvalues, (a) the lowest eigenvalue
(b) the first excited state (c) the second excited state (d) the third excited state.
The volumes are (from the left) $16^4$, $12^4$, $10^4$ and $8^4$.
}
\label{fig:several4}
\end{figure}

I begin by fitting
individual eigenvalues (lowest, second, and so on) to a power law,
$\ln \langle \lambda_i \rangle = A_i - p \ln L$. I choose to fit to all four volumes,
 or the largest three. The individual data points in each fit are uncorrelated,
of course.
Fits and chi-squareds are shown in Table \ref{tab:fittable}.
Examples of fits are shown in Figs.~\ref{fig:several3} and \ref{fig:several4}.

\begin{figure}
\begin{center}
\includegraphics[width=0.7\textwidth,clip]{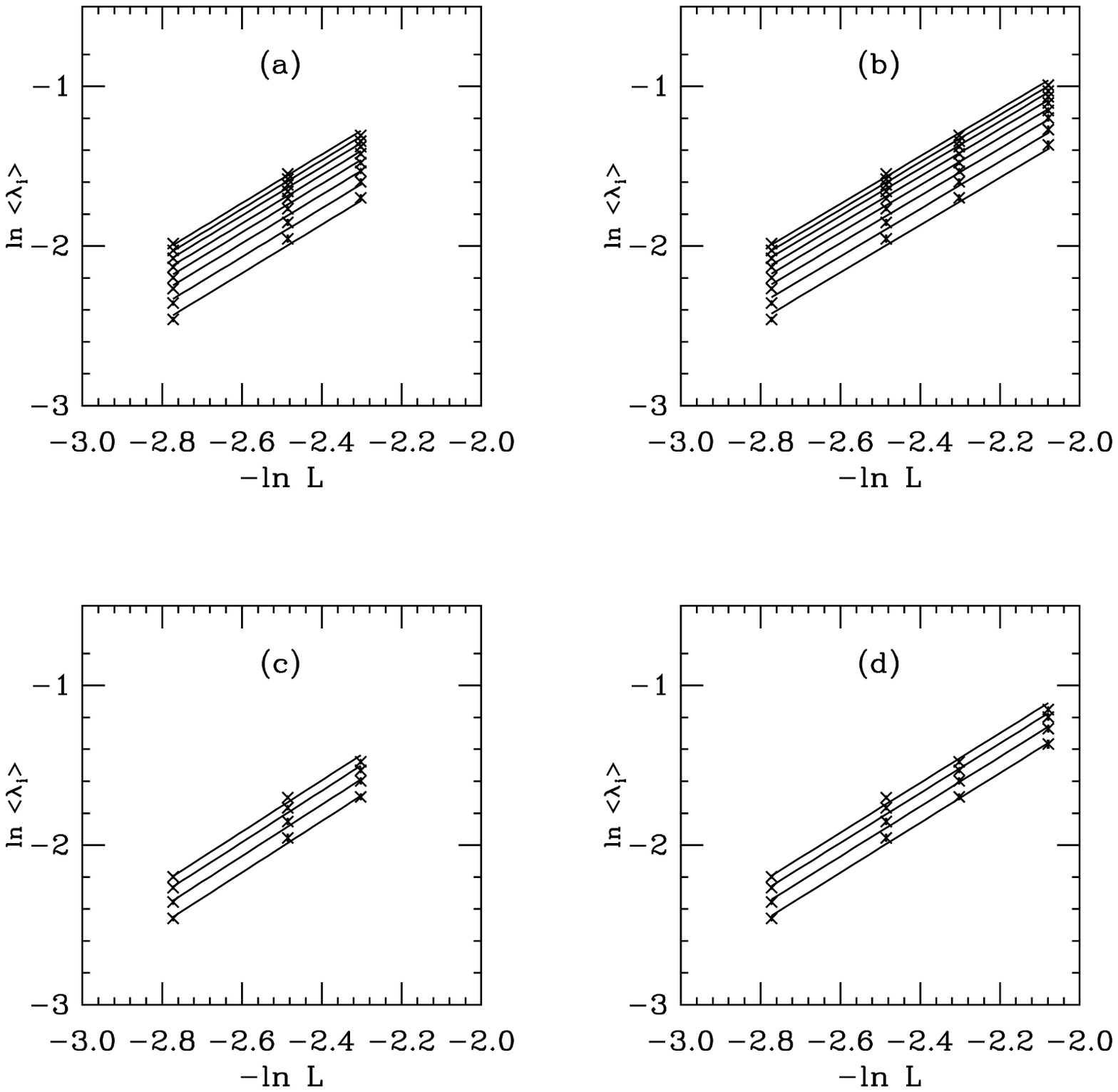}
\end{center}
\caption{Combined fits to several eigenvalues.
(a) Three-volume fits to the 8 lowest eigenvalues, (b) four volume fits to the lowest 8
eigenvalues,
(c) Three-volume fits to the  lowest 4 eigenvalues, (d) four volume fits to the lowest 4
eigenvalues.
The volumes are (from the left) $16^4$,  $12^4$ $10^4$ and in (b) and (d)
$8^4$.
}
\label{fig:several}
\end{figure}

I also fit groups of eigenvalues, using
$\ln \langle \lambda_i \rangle = A_i - p \ln L$ for $i=1\dots N$ eigenvalues.
Now the data are correlated. I look at the quality of fits from uncorrelated
fits, and then repeat by taking bootstrap averages of the data.
The behavior is quite similar to the fits to individual eigenvalues. Some examples
(with $2\sigma$ bootstrap errors) on the average, chi-squared from uncorrelated fits:
\begin{itemize}
\item Fit the lowest 4 eigenvalues and biggest 3 volumes: $p=1.59(5)$,
$\chi^2/dof=40/(12-5)$
\item Fit the lowest 4 eigenvalues 4 volumes: $p=1.54(4)$, $\chi^2/dof=61/(16-5)$
\item Fit all 8 eigenvalues and biggest 3 volumes: $p=1.51(4)$, $\chi^2/dof=126/(24-9)$
\item Fit all 8 eigenvalues and 4  volumes: $p=1.47(3)$, $\chi^2/dof=180/(32-9)$
\end{itemize}
Examples of these fits are shown in Fig. \ref{fig:several}.

\begin{table}
\caption{Exponent $p$ from fits to individual eigenvalues, from the largest three
volumes ($16^4$,  $12^4$, $10^4$), or
or to all volumes (add $8^4$).}
\begin{center}
\begin{tabular}{c|cc|cc}
\hline\hline
  &  3 volumes & & 4 volumes &  \\
mode &  $p$ &  $\chi^2$ &  $p$ &  $\chi^2$  \\
\hline
1 & 1.64(4) & 3.0  & 1.60(3) & 5.2 \\
2 & 1.64(4) & 4.3  & 1.59(3) & 8.0 \\
3 & 1.61(3) & 11.4 & 1.56(2) & 16.5 \\
4 & 1.58(3) & 18.7 & 1.52(2) & 26.2 \\
5 & 1.53(2) & 13.7 & 1.49(2) & 19.2 \\
6 & 1.51(3) & 10.0 & 1.47(2) & 17.0 \\
7 & 1.48(2) & 11.6 & 1.45(2) & 17.0 \\
8 & 1.45(2) & 8.4  & 1.43(2) & 11.4 \\
\hline\hline
\end{tabular}
\end{center}
\label{tab:fittable}
\end{table}

\begin{figure}
\begin{center}
\includegraphics[width=0.7\textwidth,clip]{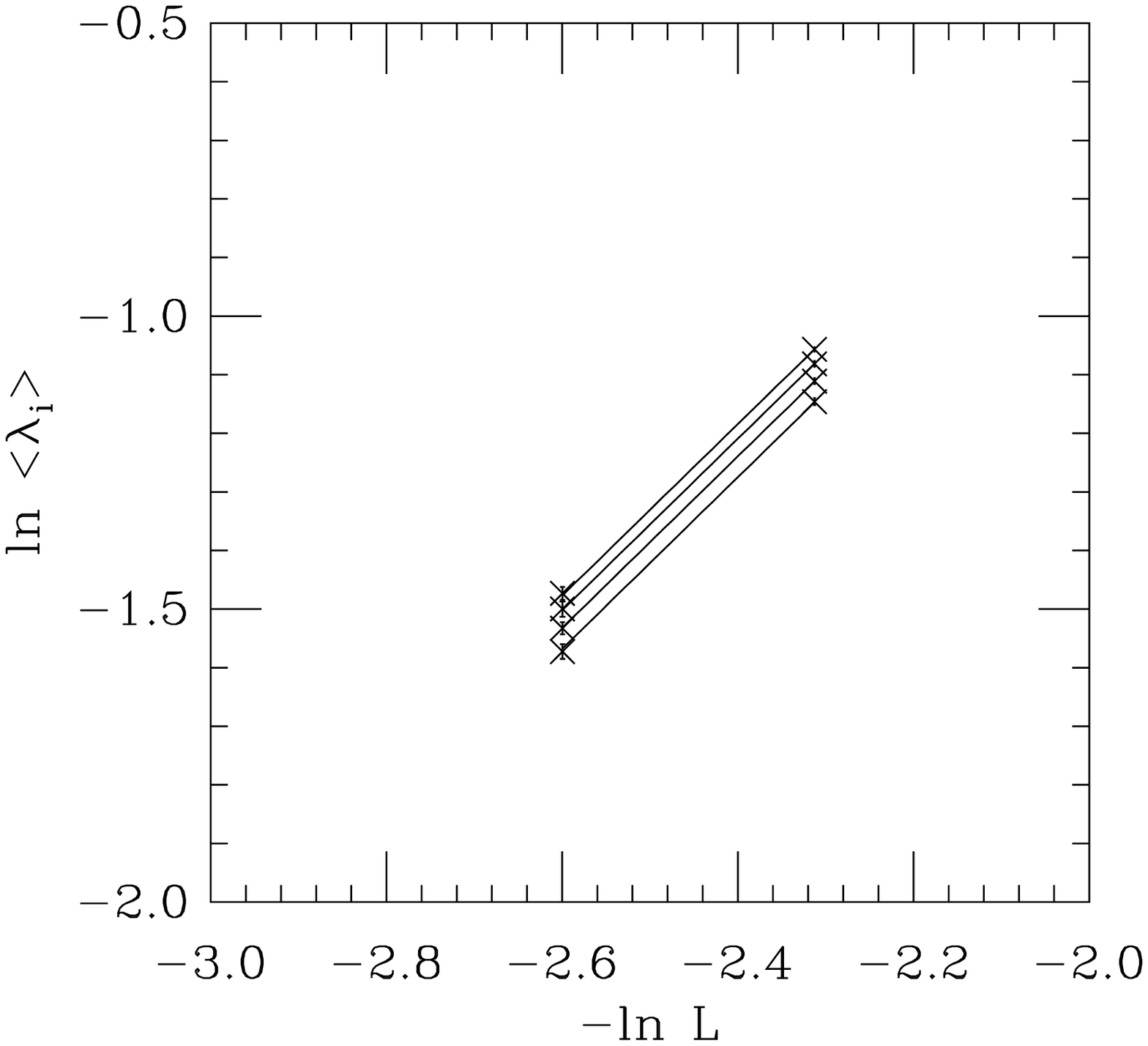}
\end{center}
\caption{Average values of 4 lowest eigenvalues of the sextet-representation
valence overlap operator, vs $1/L$, for $12^3\times 6$ and $16^3\times 8$ volumes.
As before, $L^4=V$ the lattice volume.
 The fit is to a common exponent,
$p=1.44(5)$.
}
\label{fig:fit24}
\end{figure}

The numerical value of the eigenvalues depends on the geometry of the lattice,
but the scaling exponent does not seem to do so. I have data from $12^3\times 6$ and
$16^3\times 8$ lattices. Fitting them as was done for the other data sets produces a
similar exponent (similar results for individual or multiple eigenmodes, from a
fit to the lowest four modes, $p=1.44(5)$). See Fig.~\ref{fig:fit24} for an example.
With two volumes, of course, one can assume any scaling law that one wants.
Nevertheless forcing a power law yields an  exponent which
 is similar to that from $L^4$ volumes.

At the end of this analysis I have many numbers, all rather similar but none
really identical. What is missing from this section is some way of estimating
the correction to scaling due to cutoff effects. In RMT, the larger eigenmodes
are the ones which are most affected by non-infrared physics. Kovacs \cite{Kovacs:2009zj}
has pointed out that in distributions like $\rho \sim \lambda^\alpha$, the lowest eigenvalue
has the broadest distribution and is most susceptible to finite statistics. 
Theoretical guidance is needed.
Nevertheless,
the observed exponents seem to be in quite good accord with the value of $y_m$ observed in
Sec.~\ref{sec:three}, $y_m \sim 1.5$ at $\beta=5.2$.

\section{Discussion}
The lattice-regulated
system of $N_f=2$ flavors of sextet fermions
coupled to $SU(3)$ gauge fields  has a weak coupling phase which is chirally restored. Within that
phase, hadronic correlation lengths (masses) show a weak dependence on the value of the bare
gauge coupling. At one observation point in this phase, I find that
 the valence quark condensate vanishes
as its (valence) mass is taken to zero.

Motivated by the realization that slow running is very similar to no running
(recall Eq.~\ref{eq:gammaruns2}),
I analyzed the correlation length in the weak coupling phase as if its massless limit were critical,
with the size of the relevant perturbation given by the AWI quark mass.
I observed the collapse of correlation length data from many lattice sizes onto a
common scaling curve. I did this at two bare parameter values. The observed exponent was
identical within rather large errors. Roughly the same exponent governs the scaling of the values
of low
lying valence overlap Dirac eigenvalues at one of the couplings.
My data cannot show whether the theory truly has only one relevant operator (with coupling
proportional to the quark mass) and the gauge coupling is irrelevant,
 or if the gauge coupling is also running very slowly.
This remains an open question.

 My analysis assumes that the correlation length
would diverge at zero AWI quark mass in the infinite volume limit.
The exponent $y_m$ is not unity, the value expected for a free field theory.
The system thus has interacting dynamics.

Recall that $y_m = 1 - \gamma_{\bar\psi \psi}$. The perturbative expectation
is $\gamma_{\bar\psi \psi}= -6 C_2(R) g^2/(16\pi^2)$. Inserting the Schrodinger functional
coupling quoted in the Introduction for these bare coupling values, perturbation theory
predicts $\gamma_{\bar\psi \psi}= -0.30$ and -0.41 at the two couplings. This is
a bit smaller than what I measure. (Of course, one could argue about the choice of a
coupling in a perturbative formula.) 

Ryttov and Sannino \cite{Ryttov:2007cx}
have a supersymmetric QCD - inspired beta function for gauge theories
with higher dimensional representations of fermions. The anomalous dimension is predicted to be
\bee
\gamma_{\bar\psi \psi} =  - \frac{11C_2(G)-4T(R)N_f}{2T(R)N_f} =  - \frac{13}{10}
\ee
for the case studied here.
My result disagrees with their prediction.

This research leaves some obvious open questions.
First, I have nothing to say about the strong coupling end of the weak coupling phase.

Next, it is unknown if this theory has an IRFP or merely runs very slowly.
Once that is known, results from this paper can be used for physics analyses.
Even with the present statistical significance of my result, there are some rather definite
conclusions
which can be drawn once the answer to this question is known.

If this theory does not have an IRFP, $y_m(g)$ can be used to run the condensate down
according to Eq.~\ref{eq:gammaruns}. $y_m$ seems to be a bit larger than perturbation theory
would give. Of course, collecting more points will give a nonperturbative determination of $y_m$
as a function (if any) of $g^2$.

If this theory is in fact an IRFP theory, my result indicates that it might not be a very
spectacular IRFP theory, and it may not be phenomenologically interesting.
This is because the exponent $y_m$ is small.
Unitarity bounds  for
 conformal field theories\cite{Mack:1976pa,Grinstein:2008qk} constrain the scaling dimension of
the leading scalar operator (which I am inferring will be the condensate, or at least what the
fermions mass couples to)
to lie in the range $3>d=4-y_m>1$. My $y_m=1.5$ is safely in that range. But in
``unparticle'' extensions of the Standard Model, new physics (NP)
at scale $M$ influences Standard Model (SM)
physics at scale $\Lambda$ through
terms in an effective Lagrangian of the form
\bee
{\cal L} = \frac{\Lambda^{d_{NP}+d_{SM}}}{M^{4+d_{NP}+d_{SM}}}{\cal O}_{SM}{\cal O}_{NP}
\ee
where the dimension of the new physics operator is $d_{NP}$ and the $\cal{O}$'s are operators
in the two sectors.
In the literature,  $d_{NP}$ is generally desired to be as small as possible
to enhance its effective coupling (the prefactor of the operator product).
Luty and Okui  \cite{Luty:2004ye}, for example, discuss models with $d$ in the range 1-2.
There is also an extensive literature relating the lower end of the conformal window
to large $y_m$, with $y_m=2$ perhaps having a connection with physics which closes the window.
 (See  
Refs.~\cite{Cohen:1988sq,Braun:2006jd,Appelquist:1996dq,Appelquist:1998rb,gardi,Kaplan:2009kr}.)
For sextet QCD, $N_f=3$ is almost certainly close to the top of
the conformal window, and $N_f=1$ is almost certainly
confined, so if $N_f=2$ is conformal with $y_m\sim 1.5$ there
 is not much of a story to be told.

There are two other cases where $y_m$ has been measured.
The first  is $SU(3)$ gauge theory with $N_f=16$ fundamental
representation flavors. Here, close to the top of the conformal window for its
representation class, Hasenfratz \cite{Hasenfratz:2009ea}
has found $y_m\sim 1$. It is not surprising that
$y_m$ is greater than 1 for $N_f=2$  sextet QCD since 
$y_m$ is expected to increase
with increasing distance from the top of the conformal window.
The authors of Ref.~\cite{Bursa:2009tj}
report a determination of $\gamma_m$ for $SU(2)$ gauge fields and $N_f=2$ adjoint
fermions. They also find it is small.

Understanding strongly-coupled beyond-Standard Model physics
requires studying many theories and computing the anomalous dimensions for enough of them
to be able to understand systematic trends.
Finite size scaling studies are a useful way to do this.

\begin{acknowledgments}
I thank P.~Damgaard, C.~DeTar, A.~Hasenfratz, A.~Jackson,   T.~G.~Kovacs, D.~N\'ogr\'adi,
S.~Pigolotti,
K.~Rummukainen,
Y.~Shamir, and B.~Svetitsky 
 for discussions.
I am grateful for the hospitality of the Niels Bohr International Academy during
the time I carried out this research.
This work was supported in part by the US Department of Energy.
%
\end{acknowledgments}

\end{document}